# Pion Fusion in Peripheral $pp$ Collisions

Alec J. Schramm [*]

*Department of Physics, Occidental College, Los Angeles, CA 90041*

Berndt Müller [†]

*Department of Physics, Duke University, Durham, NC 27708-0305*

## Abstract

We estimate the cross section for quasi-elastic double pion exchange in high energy proton-proton collisions. Total and elastic $\pi\pi$ cross sections are calculated in an equivalent pion approximation, with pion-baryon vertices taken from chiral perturbation theory.
13.75.Cs,13.75.Gx,12.39.Hg

Typeset using REVTEX

[*]Email: *alec@phys.oxy.edu*

[†]Email: *muller@phy.duke.edu*

## I. INTRODUCTION

One of the motivations for the development of a Full Acceptance Detector at hadron colliders [1,2] is to exploit the rapidity gaps which can appear in hadron-hadron collisions [3]. These gaps could serve as a clean experimental signature for both the verification of the Standard Model and the discovery of new physics beyond. For example, virtual $W$ bosons accompanying high-energy hadron beams, treated as partons of the incoming protons, can fuse to create a Higgs boson which subsequently decays into $W^+W^-$ pairs. The leptonic decay of these $W$ bosons will occur in a region of rapidity in which no hadrons are found, thus facilitating identification of the Higgs [4]. There are, of course, backgrounds to this process which need to be considered, such as strong double diffractive scattering, i.e., double Pomeron exchange.

Another opportunity afforded by a Full Acceptance Detector is the ability to measure total and elastic meson-meson cross sections, using leading particle tags to isolate meson exchange contributions. This technique would among other things allow for the study of pion-pion and pion-nucleon interactions at a considerable fraction of the total center-of-mass energy (several TeV at the CERN Large Hadron Collider). In a $pp$ or $p\overline{p}$ collider, virtual pions associated with the beams can fuse; this process can be isolated by requiring two forward-going quasi-elastically scattered $\Delta$ baryons, identified through their decay into nucleons and pions. In peripheral $pp$ collisions, this would lead to the process shown in figure 1,

$$pp \rightarrow \Delta + \Delta + \pi + \pi \rightarrow \Delta + \Delta + X$$

A possible theoretical approach to such events uses an "equivalent pion approximation", developed by analogy with the Weizsäcker-Williams method of virtual quanta in electrodynamics. This approximation is based on the idea that the meson field accompanying an energetic baryon consists of virtual quanta close to mass shell, and as such can be replaced with an equivalent pulse of real pions. For pions with low momenta, the results of heavy baryon chiral perturbation theory for low energy pion-baryon interactions can be used to determine the pion-baryon vertices. In this paper we use these methods to study the process $\pi\pi \to X$. We calculate the cross section $\sigma_{pp\to\Delta\Delta X}$ and determine its dependence on the impact parameter of the $pp$ system. The effects of absorption are taken into account within the framework of the Glauber approximation. In section 2 we recount the basics of heavy baryon chiral theories. The relevant portions are applied in section 3 to the equivalent pion approximation, and our results are discussed in section 4.

## II. BARYON CHIRAL PERTURBATION THEORY

The chiral description of hadron interactions is based upon an $SU(2)_L \times SU(2)_R$ chiral symmetry spontaneously broken to the diagonal $SU(2)$ subgroup. The pseudoscalar mesons emerge as Goldstone bosons, and are described in an effective Lagrangian as a power series expansion in derivatives. Higher dimensional operators are suppressed by powers of $m_\pi/\Lambda_\chi$, where $\Lambda_\chi \sim 1$ GeV is the chiral symmetry breaking scale. Inclusion of baryons in this scheme is achieved by treating the baryon as a heavy fermion interacting with low momentum pions. For consistency, octet as well as decuplet baryons must be included; this then yields a rapidly convergent perturbation expansion [5].

In the heavy baryon limit, the momentum transferred between baryons via pion exchange is small compared to the baryon mass; as a result, the velocity of the baryon is virtually unaffected. In fact a field with definite position and velocity can be obtained, since in the heavy baryon limit

$$[v^\mu, x^\nu] = i\hbar g^{\mu\nu}/M_B \longrightarrow 0 \ . \tag{1}$$

The fields of definite velocity $v^\mu$ are related to the original baryon fields by

$$B_V(x) = e^{iM_B \slashed{v} v_\nu x^\nu} B(x) \ , \tag{2}$$

where $M_B$ is the baryon mass; similarly, for the spin-3/2 decuplet a Rarita-Schwinger field of definite velocity can be defined by

$$T_v^\mu = e^{iM_T \slashed{v} v_\nu x^\nu} T^\mu(x) \ . \tag{3}$$

One can write an effective theory using the new baryon field $B_v$, which obeys the modified Dirac equation

$$i\slashed{\partial} B_v = 0 \ . \tag{4}$$

Note that the usual mass term does not appear in this equation. In fact by expressing the momentum of a nearly on-shell baryon as [6]

$$p^\mu = M_B v^\mu + k^\mu \ , \tag{5}$$

where $v \cdot k \ll M_B$ is proportional to the amount by which the baryon is off-shell, we see that derivatives acting on $B_v$ produce factors of $k$ rather than $p$. Thus higher derivative terms in the Lagrangian are suppressed by powers of the small number $k/\Lambda_\chi$, allowing for a systematic Lagrangian expansion in powers of derivatives.

For our purposes, we need the expression for the $\pi p \Delta$ vertex. Each pion-baryon vertex contributes a factor of [5]

$$(c_{\pi BT}) \frac{C}{\sqrt{2} F_\pi} k^\mu \ , \tag{6}$$

where $F_\pi = 93$ MeV is the pion decay constant, $k^\mu$ is the pion momentum, and $c_{\pi BT}$ is the Clebsch-Gordan coefficient at the $\pi$BT vertex. The constant $C = 1.53$ has been determined [5] by fitting the experimentally measured $T \to B\pi$ decay modes. In addition to this vertex factor, there will be a contribution from the spin sum over the $B_v$ and $T_v^\mu$ fields. For the interaction shown in figure 1 the invariant matrix element is (we omit the subscript $v$ for simplicity)

$$\mathcal{M} \sim \overline{T}_\mu(p_1') \, k_1^\mu \, B(p_1) \, A_{\pi\pi \to X}^I \, \overline{T}_\nu(p_2') \, k_2^\nu \, B(p_2) \ , \tag{7}$$

where $A_{\pi\pi \to X}^I$ is the amplitude for two pions of total isospin $I$ to fuse into anything. Averaging and summing over initial and final spins, respectively, will give at each $TB\pi$ vertex the factor

$$V_{TB\pi} = \frac{1}{2}\sum_{s,s'}\overline{T}_\mu(p',s')\, k^\mu\, B(p,s)\, \overline{B}(p,s)\, \tilde{k}^\nu\, T_\nu(p',s') \,. \tag{8}$$

In order to utilize the normal trace techniques, we need to write the completeness relations for spinor and Rarita-Schwinger fields in terms of the heavy-baryon velocity,

$$\sum B\overline{B} = (M_B + \not{p}) = M_B(1 + \not{v}) \tag{9}$$

$$\sum T_\mu \overline{T}_\nu = (M_T + \not{p})P_{\mu\nu} = M_T(1+\not{v})P_{\mu\nu} \,. \tag{10}$$

In the latter expression

$$P^{\mu\nu} = v^\mu v^\nu - g^{\mu\nu} - \frac{4}{3}S^\mu S^\nu \,, \tag{11}$$

where $S^\mu$ is the spin operator acting on the spinor indices of $T^\mu$ [5], satisfying

$$\{S^\mu, S^\nu\} = \frac{1}{2}(v^\mu v^\nu - g^{\mu\nu}) \,. \tag{12}$$

With these in hand, we find

$$V_{TB\pi} = \frac{4}{3}M_T M_B \left(k\cdot v'\,\tilde{k}\cdot v' - k\cdot\tilde{k}\right)(1 + v\cdot v') \,. \tag{13}$$

If we let $x$ denote the fraction of the proton energy carried off by the pion,

$$xE = E - E' \,, \tag{14}$$

then with $\vec{p}'_\perp = 0$ in the heavy baryon limit we find

$$V_{TB\pi} = \frac{2}{3}(T+B)^2 \left[\frac{1}{4M_T^2}\left(T^2 - B^2\right)^2 + \vec{k}_\perp \cdot \vec{\tilde{k}}_\perp\right] \,, \tag{15}$$

where the scalar functions $T$ and $B$ are defined as

$$T \equiv M_T/\sqrt{1-x} \tag{16}$$

$$B \equiv M_B\sqrt{1-x} \,. \tag{17}$$

## III. EQUIVALENT PION APPROXIMATION

The equivalent pion approximation is the generalization of the equivalent photon approximation used in earlier work [7,8]. The idea is that the meson field of a fast baryon can be replaced with an equivalent pulse of real pions. For proton beams $A$ and $B$ with center of mass energy $s$, one assumes that the cross section for $\pi\pi$ fusion can be written in the form

$$\sigma_{AB}^{\pi\pi} = \int dx_1 dx_2 \, f_\pi^A(x_1) f_\pi^B(x_2) \sigma_{\pi\pi}(s_{\pi\pi}) \,, \tag{18}$$

where $x_i$ is the fractional beam momentum carried by the virtual pions, $\sigma_{\pi\pi}$ is the production cross section in the collision of two real pions with squared center of mass energy $s_{\pi\pi} \equiv x_1 x_2 s$, and $f_\pi(x)$ is the quasi-elastic pion distribution function of the proton. Using the results from the previous section, we find

$$f_\pi(x) = G \, x(T+B)^2 \int_0^\infty \frac{d^2 k_\perp}{(2\pi)^2} \left[ \frac{(T^2 - B^2)^2}{4 M_T^2} + k_\perp^2 \right] \frac{F(-k^2)^2}{(k^2 - m_\pi^2)^2} \,. \tag{19}$$

In this expression

$$-k^2 = \frac{x^2 M_B^2 + x(M_T^2 - M_B^2)}{1-x} + k_\perp^2 \,, \tag{20}$$

$F(-k^2)$ is the $\pi N \Delta$ form factor, and

$$G \equiv \frac{c_{\pi\pi X} \, c_{\pi BT}^2 C^2}{12 \pi F_\pi^2} \,. \tag{21}$$

In the spirit of the eikonal approximation, we write the total $\pi\pi$ exchange cross section (18) in the impact parameter representation and fold it with the probability that no inelastic interaction takes place other than double pion exchange. The dependence of the cross section on impact parameter is then found by integrating the squared matrix element over all spacetime coordinates save for the transverse distance $b$ between the protons. Following the procedure in [8], the differential cross section is

$$\frac{d^2 \sigma_{AB}^{\pi\pi}}{d^2 b} = \int_0^1 dx_1 \int_0^1 dx_2 \int \frac{d^2 q}{(2\pi)^2} \, e^{i\vec{q}\cdot\vec{b}} \, f_\pi^A(x_1, q) f_\pi^B(x_2, q) \, \sigma_{\pi\pi}(x_1 x_2 s) \tag{22}$$

where

$$f_\pi(x,q) = G\ x(T+B)^2 \int_0^\infty \frac{d^2k_\perp}{(2\pi)^2} \left[\frac{(T^2-B^2)^2}{4M_T^2} + (k_\perp^2 - q^2/4)\right]$$

$$\times \frac{F[-(k+q/2)^2]}{[(k+q/2)^2 - m_\pi^2]} \frac{F[-(k-q/2)^2]}{[(k-q/2)^2 - m_\pi^2]}, \quad (23)$$

and $-k^2$ is given by equation (20).

## IV. RESULTS

We studied the specific process $pp \to \Delta^{++} + \Delta^{++} + X$, requiring the fusion of two $\pi^-$ mesons; this reaction can be triggered on the four positively charged decay products of the $\Delta^{++}$ baryons. For this process, the Clebsch-Gordan coefficient is $c_{\pi BT} = c_{\pi N\Delta} = 1/\sqrt{2}$.

In the calculations, we used the standard dipole form factor of the nucleon,

$$F(\vec{k}^2) = \left[1 + (\vec{k}^2/.71\text{GeV}^2)\right]^{-2}. \quad (24)$$

This form factor is in quite good agreement with the measurements of the form factor of the delta-pion-nucleon vertex at small momenta [9], where it is needed in our calculations. In order to obtain a quantitative estimate, the cross section $\sigma_{\pi\pi}(x_1 x_2 s)$ was taken to be 4/9 times the nucleon-nucleon cross section $\sigma_{NN}$, assuming the validity of additive quark model counting rules. With $\sigma_{NN} \approx 70$ mb at 1 TeV [10], our calculation yields a total cross section of $\sigma(pp \to \Delta^{++}\Delta^{++}X) = 25\mu$b for 1 TeV protons. Of course inelastic scattering effects due to the $\Delta$ baryons will reduce this somewhat. This decrease was accounted for using the Glauber approximation by an absorption factor [11]

$$\frac{d\sigma_{NN}^{\text{el}}}{d^2 b} = \frac{d\sigma_{NN}}{d^2 b}\ \exp\left[-\sigma_0 T_{NN}(b)\right], \quad (25)$$

where $\sigma_0$ is the total nucleon-nucleon cross section. For simplicity, we have used a gaussian approximation to the form factor (24) in calculating the profile function [12]

$$T_{NN}(b) = \int \frac{d^2 Q}{(2\pi)^2}\ F_N(Q^2)\ F_N(Q^2)\ e^{iQb}. \quad (26)$$

Including these absorption effects reduces the total double-pion exchange cross section to about 9.7 $\mu$b. The majority of the inelastic events is expected to occur at small values of the impact parameter $b$. Indeed, the quasi-elastic nature of the interaction is maintained only in those collisions in which the two protons pass by each other. Thus it is important to verify that a significant portion of the $\pi\pi$ cross section extends out to relatively large impacts parameters. Note that since we are only concerned with this high-$b$, low $k_\perp$ region, higher order $\pi N \Delta$ vertices do not need to be considered. The dependence of the differential cross section on impact parameter is shown in figure 2, where the dotted curve shows the suppressing effects of absorption.

As may be expected, the process $pp \to \Delta\Delta X$ also receives contributions from $\pi\rho$ and $\rho\rho$ fusion. We have estimated these modes using distribution functions derived from a convolution model. In this approach, the nucleon is expanded as a sum of Fock states consisting of a virtual meson $M$ and a virtual baryon $B$ with longitudinal momentum fractions $x$ and $1-x$, respectively. For our calculation, the distribution function is given by a sum of contributios from all possible $\rho\Delta$ helicity states [13]. As form factor, we use

$$F_{\rho\Delta}(x, k_\perp^2) = \exp\left[-\frac{M_{\rho\Delta}^2(x, k_\perp^2) - m_N^2}{2\Lambda^2}\right] \qquad (27)$$

where $M_{\rho\Delta}^2(x, k_\perp^2)$ is the invariant mass squared of the $\rho\Delta$ state

$$M_{\rho\Delta}^2(x, k_\perp^2) = \frac{m_\rho^2 + k_\perp^2}{x} + \frac{m_\Delta^2 + k_\perp^2}{1-x} \; ; \qquad (28)$$

data fitting gives $\Lambda = 0.98$ [13].

Taking the $N\Delta\rho$ coupling to be $f^2/4\pi = 34.7$ Gev$^{-2}$ [14], and estimating $\sigma_{\rho\rho} = \sigma_{\pi\rho} = \sigma_{\pi\pi}$, we find that $\rho\rho$ fusion adds only 9.4 $\mu$b to the total cross section, while $\pi\rho$ fusion contributes 15.4 $\mu$b. Due to the greater mass of the $\rho$, these processes are not expected to contribute as much to the total cross section—and in fact will suffer greater absorptive suppression than will $\pi\pi$ fusion. Indeed, absorption reduces the $\pi\rho$ contribution to 5.7 $\mu$b; figure 3 compares of the $\pi\pi$ and $\pi\rho$ impact parameter dependences.

Of some interest is the cross section as a function of the rapidity of the two pion state. Neglecting the transverse momenta of the equivalent pions, we can write the rapidity as

$$y = \frac{1}{2} \ln \frac{\epsilon + p}{\epsilon - p} = \frac{1}{2} \ln \frac{x_1}{x_2} , \qquad (29)$$

where $\epsilon \equiv (x_1 + x_2)E$ is the total energy of the $\pi\pi$ state, and $p \equiv (x_1 - x_2)E$ is its momentum. We can then calculate the differential cross section $d\sigma/dy$ from equation (18), using the variables $u = x_1 x_2$ and $y = \frac{1}{2} \ln(x_1/x_2)$. The rapidity dependence of the cross section (without the effects of nuclear absorption) is shown in figure 3. Given that the range of rapidity for the nucleon projectiles is roughly between $\pm 8$, a virtual pion rapidity of $|y| < 2$ seems reasonable. Figure 4 shows the invariant mass spectrum, both with and without absorption. The cross section peaks at an invariant mass of $M_X \approx 60$ MeV; absorption effects reduce this contribution by about a factor of 2.5.

## V. CONCLUSIONS

We have estimated the rate of particle production via fusion of virtual pions in peripheral $pp$ collisions. By analogy with equivalent photon techniques, we have developed an equivalent pion approximation; pion-baryon vertices were determined using the results of heavy baryon chiral perturbation theory for low energy pion-baryon interactions. Contributions from $\rho$ fusion were also estimated using the results of a convolution model. Absorption considerations were accounted for using the Glauber approximation. For the representative process $pp \to \Delta^{++} + \Delta^{++} + X$ at 1 TeV, we estimate that a luminosity of $10^{31} \text{s}^{-1} \text{cm}^{-2}$ would yield about 150 events per second due to $\pi\pi$ and $\pi\rho$ fusion; an integrated luminosity of 10 pb$^{-1}$ produces some 150 million events.

## ACKNOWLEDGMENTS

We thank Professors R.P. Springer and W.D. Walker for useful discussions. One of us (AJS) would like to thank Occidental College for travel support and Duke University for support and hospitality while part of this work was done. This research was partially supported by DOE under grant DE-FG05-90ER40592.

# REFERENCES

[1] J.D. Bjorken, SLAC preprint SLAC-PUB-5927 (1992); SLAC preprint SLAC-PUB-6352 (1993).

[2] K.L. Kowalski and C.C. Taylor, Case Western preprint CWRUTH-92-6 (1992); J. D. Bjorken, K.L. Kowalski and C.C. Taylor, "Observing Disoriented Chiral Condensates", Proceedings of the Workshop on *Physics at Current Accelerators and Supercolliders*, Argonne National Laboratory June 2-5 1993, Edited by J.L. Hewett, A.R. White and D. Zeppenfeld, pp. 73-88.

[3] E. Levin, Phys. Rev. **D48** (1993), 2097; E. Gotsman, E.M. Levin, and U. Maor, Phys. Lett. **B309** (1993), 199.

[4] Y. Dokshitzer, V. Khoze, and S. Troyan, *Proceedings of the Sixth Annual Conference of Physics in Collisions* (1986), ed. M. Derrick (World Scientific, Singapore, 1987), 365; J.D. Bjorken, Phys. Rev. **D47** (1993), 101.

[5] E. Jenkins and A. Manohar, Phys. Lett. **255B** (1991), 558; *Effective Field Theories of the Standard Model*, Dobogókő, Hungary, edited by U. Meissner, World Scientific (1992), 113.

[6] H. Georgi, Phys. Lett. **240B** (1990), 447.

[7] E. Fermi, Z. Phys. **29** (1924), 315; E.J. Williams, Proc. Roy. Soc. **A139** (1933), 163; C. Weizsäker, Z. Phys. **88** (1934), 612.

[8] B. Müller and A.J. Schramm, Phys. Rev. **D42** (1990), 3699.

[9] P. Stoler, Phys. Rev. Lett. **66** (1991), 1003; Phys. Rev. **D44** (1991), 73.

[10] Particle Data Group, "Review of Particle Properties", Phys. Rev. **D50** (1994) Part I.

[11] V. Franco and R.J. Glauber, Phys. Rev. **142** (1966), 1195.

[12] T.T. Chou and C.N. Yang, Phys. Rev. **170** (1968), 1591.

[13] H. Holtmann, A. Szczurek, and J. Speth, Jülich preprint KFA-IKP(TH)-1993-33, and references therein.

[14] B. Holzenkamp, K. Holinde, and J. Speth, Nuc. Phys. **A500** (1989), 485.

**FIGURES**

FIG. 1. The fusion of two virtual pions from scattering protons.

FIG. 2. The impact parameter dependence of the differential cross section for $\pi^-\pi^- \to X$ in the scattering of 1 TeV protons. The solid curve is without absorption; the dotted curve includes absorption effects.

FIG. 3. Comparison of the the differential cross section for $\pi\pi \to X$ with $\pi\rho \to X$. The solid and dashed curves are the same as in Fig. 2; the dotted-dashed and dotted curves are the respective analogs for $\pi\rho$ fusion.

FIG. 4. The dependence of the differential cross section on rapidity $y$.

FIG. 5. The dependence of the differential cross section on invariant mass $M_X$. The solid curve is without absorption; the dotted curve includes absorption effects.

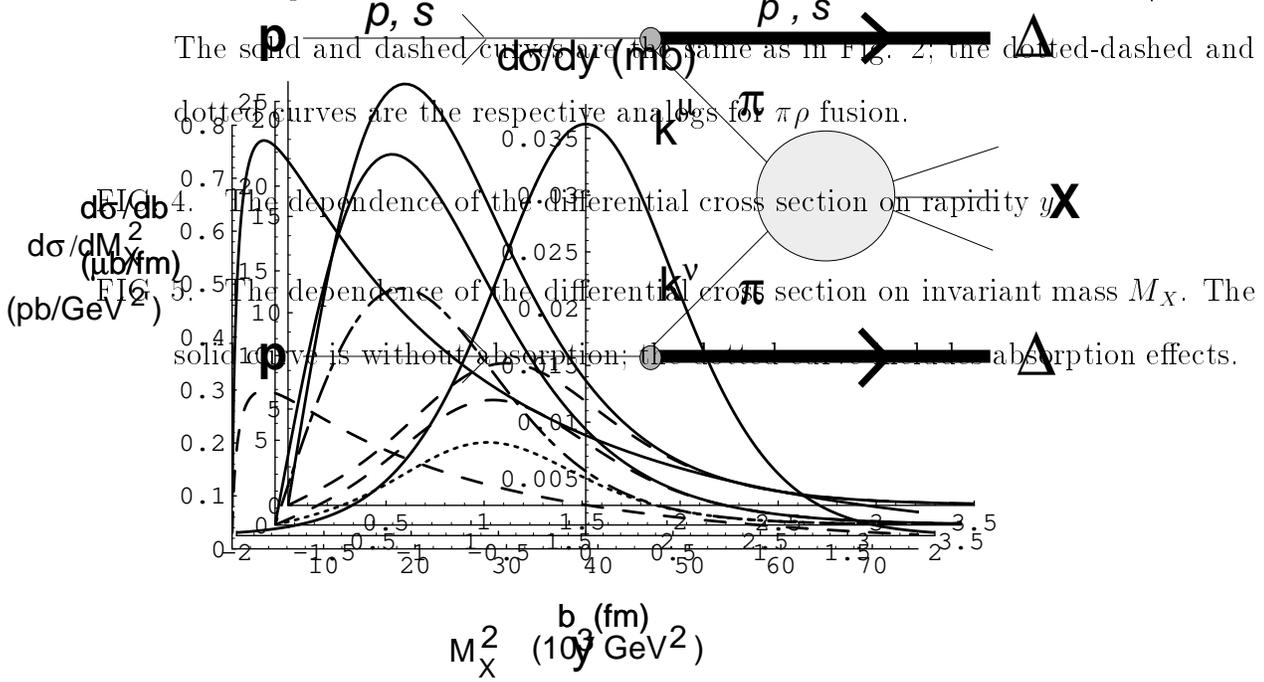